\documentclass[conference]{IEEEtran}
\IEEEoverridecommandlockouts
\usepackage{fancyhdr}
\usepackage{cite}
\usepackage{amsmath,amssymb,amsfonts}
\usepackage{algorithmic}
\usepackage{graphicx}
  \graphicspath{{Figures/}}
\usepackage{textcomp}
\usepackage{xcolor}
\usepackage[a4paper, total={184mm,239mm}]{geometry}
\def\BibTeX{{\rm B\kern-.05em{\sc i\kern-.025em b}\kern-.08em
    T\kern-.1667em\lower.7ex\hbox{E}\kern-.125emX}}

\usepackage[USenglish]{babel} 
\usepackage{url}
\usepackage{hyperref}
\usepackage[utf8]{inputenc}
\usepackage[shortcuts]{extdash}
\usepackage{xfrac}
\usepackage{verbatim}
\usepackage{array}

\usepackage{cleveref}
\crefname{section}{Section}{Sections}
\crefname{appendix}{Section}{Sections}
\crefname{figure}{Fig.}{Fig.}
\Crefname{figure}{Figure}{Figures}
\crefname{table}{Table}{Tables}
\crefname{equation}{Eq.}{Eq.}
\Crefname{equation}{Equation}{Equations}

\usepackage{color}
\usepackage{soul}

\usepackage{xspace}

\begin{document}

\title{Resilient Endurance-Aware NVM-based PUF against Learning-based Attacks\IEEEauthorrefmark{4}

\thanks{\IEEEauthorrefmark{4}This is the authors version of the extended abstract to appear at DATE\,2025.\\
This work is supported in part by the German Research Foundation (DFG) as part of the priority program “SPP 2377: Disruptive Memory Technologies” under project:  (ARTS-NVM) and in part by the German Federal Ministry of Education and Research (BMBF) through grant 01IS23066 as part of the Software Campus Project ``HE-Trust''. We thank DFG (Project Number: 405422836, NVM-OMA).}
}
\author{\IEEEauthorblockN{Hassan Nassar\IEEEauthorrefmark{1}, Ming-Liang Wei\IEEEauthorrefmark{2}, Chia-Lin Yang\IEEEauthorrefmark{2}, Jörg Henkel\IEEEauthorrefmark{1}, Kuan-Hsun Chen\IEEEauthorrefmark{3}}\\

\IEEEauthorblockA{\IEEEauthorrefmark{1}\textit{Karlsruhe Institute of Technology (KIT), Chair for Embedded Systems (CES), Germany} \\
\IEEEauthorrefmark{2}\textit{National Taiwan University, Taiwan}
\IEEEauthorrefmark{3}\textit{University of Twente, the Netherlands} \\
\IEEEauthorrefmark{1}\{nassar, henkel\}@kit.edu, \IEEEauthorrefmark{2}d04943004@ntu.edu.tw, yangc@csie.ntu.edu.tw
\IEEEauthorrefmark{3}k.h.chen@utwente.nl}  
}

\maketitle
\renewcommand{\headrulewidth}{0.0pt}
\thispagestyle{fancy}
\lhead{}
\rhead{}
\chead{This is the author's version of the work.
The definitive Version of Record (two page extended abstract) will appear in the 2025 ACM/IEEE Design, Automation and Test in Europe Conference (DATE).}
\cfoot{}
\begin{abstract}
Physical Unclonable Functions (PUFs) based on Non-Volatile Memory (NVM) technology have emerged as a promising solution for secure authentication and cryptographic applications. 
By leveraging the multi-level cell (MLC) characteristic of NVMs, these PUFs can generate a wide range of unique responses, enhancing their resilience to machine learning (ML) modeling attacks. 
However, a significant issue with NVM-based PUFs is their endurance problem; frequent write operations lead to wear and degradation over time, reducing the reliability and lifespan of the PUF. 

This paper addresses these issues by offering a comprehensive model to predict and analyze the effects of endurance changes on NVM PUFs. 
This model provides insights into how wear impacts the PUF's quality and helps in designing more robust PUFs.
Building on this model, we present a novel design for NVM PUFs that significantly improves endurance. 
Our design approach incorporates advanced techniques to distribute write operations more evenly and reduce stress on individual cells. 
The result is an NVM PUF that demonstrates a $62\times$ improvement in endurance compared to current state-of-the-art solutions while maintaining protection against learning-based attacks. 
\end{abstract}


\section{Introduction}
\label{sec:intro}

Physical Unclonable Functions (PUFs) have become crucial in modern security systems by generating unique identifiers through the intrinsic physical variations present in each device~\cite{capuf}.
PUFs use a challenge-response protocol.
It receives an input \textit{`challenge'} to produce an output \textit{`response'}.
The deployment of PUFs in the field is preceded by an enrollment phase, where a trusted third party gives the PUF challenges to collect challenge response pairs (CRPs), which will be used later to authenticate the PUF.
The physical variations cause unique, nearly random responses, even with identical challenges to PUFs with the exact same design. Thus, the behavior of a PUF cannot be cloned from another identical PUF. This inherent unclonability makes PUFs highly valuable for authentication, key storage, and secure communications, safeguarding sensitive data even in potentially hostile environments~\cite{IPA20}.

However, recent works show that PUFs are vulnerable to learning-based attacks~\cite{inter_attack,attackXOR15}. Attackers who can collect sufficient CRPs from a PUF can train ML models to predict responses to unseen challenges. This ability undermines the security guarantees provided by PUFs, as the attacker can effectively bypass the device's unique challenge-response relationship by approximating it with a learning model. Consequently, ensuring that PUFs remain resilient to such attacks is a pressing concern.
Unlike trusted third-party enrollment, attackers collect CRPs by unintendedly monitoring them, e.g., using man-in-the-middle attacks, during deployment.

Recent works leverage Non-Volatile Memory (NVM) technologies as a potential mitigation against learning-based attacks. NVM-based PUFs utilize the unique characteristics of NVM cells, such as their ability to gradually change states with iterative pulsing. This gradual change introduces a level of complexity that can obscure the cell's behavior from ML models, thereby enhancing resilience to prediction attacks~\cite{anv-puf,Arbiter_mpuf}. The iterative pulsing of NVM cells can generate a diverse set of responses, which serve as a powerful countermeasure against ML-based modeling.

Despite their benefits, NVM technologies face endurance problems due to repeated writes. Quality and reliability degrade over time with multiple write cycles, affecting longevity and consistency. Solutions such as wear leveling have been proposed to balance writes when NVM is used as main memory~\cite{nvmWear23, DBLP:journals/tecs/HakertCSBGBAHC22}. 

However, in the context of PUFs, the endurance issue is rarely discussed, and it is crucial to the applicability of such PUFs. 
In particular, while wear leveling will ensure uniform endurance degradation of the cells, it will not stop the change in the behavior of individual cells.
This behavior is crucial for the reproducibility of the responses.
Any behavior change causes PUF noise, making deployment responses differ from enrollment responses, leading to authentication failures as the response value will not match the expected response.
Hence, addressing this endurance problem is critical for the practical deployment of NVM-based PUFs in real-world applications.

\noindent\textbf{Our Contributions:} 
Most works on NVM-based PUFs consider general solutions for delay-based silicon PUFs that would work for NVM-PUFs.
However, since these PUFs are significantly different, such an assumption must be examined.
To the best of our knowledge, this is the first work to address the endurance of NVM-based PUFs.
To address this pressing issue, in a nutshell, we provide the following contributions:
\begin{itemize}
    \item We provide an analytical model for the endurance of NVM-based PUFs, analyzing how repeated writes impact their quality and reliability. Several state-of-the-art PUF designs are investigated and their susceptibility to endurance degradation are reported.
    \item We propose a novel endurance-aware PUF design, named REAP-NVM, for an NVM-based PUF that mitigates learning-based attacks while addressing the endurance issues of NVM-based PUFs.
    \item We evaluate the lifetime, energy consumption, and performance with REAP-NVM. Specifically, the results show that compared with the baseline, our design achieves 62$\times$ improvement in the endurance.
\end{itemize}

The remainder of the paper is structured as follows. \Cref{sec:Back} gives the necessary background on PUFs and NVM endurance.
We show our endurance analysis framework in \cref{sec:EndurNVM} and propose an endurance-aware NVM-PUF in \cref{sec:proTech}.
We evaluate our design in contrast to the state-of-the-art in \cref{sec:res} and provide conclusions in \cref{sec:conc}.

\section{Background}
\label{sec:Back}
PUFs are classified into weak and strong types~\cite{puf_fpga_esl}. 
Weak PUFs have limited responses, suitable for key storage. Strong PUFs offer more responses, ideal for secure authentication and key generation. 
Two main technologies are delay-based silicon PUFs, which measure signal delays, and NVM-based PUFs, the focus of this work, which exploit memory cell state unpredictability to produce unique responses.

\subsection{Quality Metrics of PUFs}
Reliability is a key metric for evaluating PUFs, especially in the context of endurance degradation. 
The other metrics are uniformity and uniqueness, each with an ideal value. 
Metrics close to the ideal indicate good performance; otherwise, performance is poor. 
Reliability, ideally at 100\%, is evaluated by  collecting the response to a challenge several times under different conditions as \cref{eq:reli} shows, where
HD is the hamming distance, $N$ is the number of times the evaluation is done, $m$ is the number of response bits, $R_s$ is the stable reference response, and $R_i$ is the response collected at one iteration out of $N$:

\begin{equation}
\label{eq:reli}
\text{Reliability}= (1-\frac{1}{N}\sum_{i=0}^{N-1} \frac{ {HD (R_{s},R_{i})}} {m})\times 100 \% 
\end{equation}

The uniformity metric measures the Hamming weight (frequency of 1s) in the response. 
Ideally, the probability of 0 and 1 should be equal, making the uniformity 50\%. The uniformity is calculated by \cref{eq:uniform}, which calculates the hamming weight of the response $R$ that has $m$ bits:

\begin{equation}
\label{eq:uniform}
\text{Uniformity}=\frac{1}{m}\sum_{i=0}^{m-1}R(i)\times 100 \% 
\end{equation}

The uniqueness metric measures how distinct a PUF is. 
Similar responses across ICs suggest that the design is governed by delay paths rather than process variations. The ideal uniqueness is 50\%. 
Higher uniqueness shows similar responses, while lower uniqueness indicates bit inversion. 
It is calculated by \cref{eq:uniq}. It is calculated on $N$ PUFs for one challenge comparing the hamming distance between responses ($R_i$ and $R_j$) of the different PUFs and normalized to the number of bits $m$.

\begin{equation} 
\label{eq:uniq}
\text{Uniqueness}=\frac{2}{N(N-1)}\sum_{i=0}^{N-2}\sum_{j=i+1}^{N-1}\frac{HD(R_{i},R_{j})}{m}\times 100 \%
\end{equation}

\subsection{Learning-based Attacks and Mitigations}

Previous works show that attackers use ML models to predict PUF behavior, compromising their unpredictability~\cite{attackXOR15}. By training on extensive PUF response datasets, adversaries can discern patterns and predict outputs accurately. To counter ML attacks, ML-resilient PUFs (that resist learning-based attacks) have been developed~\cite{ctpuf20}. PUFs integrating cryptographic functions make the prediction of responses much harder~\cite{lpn17}. But this comes with a significant overhead~\cite{capuf}.

Another direction is to use NVMs, as they are particularly suited for ML-resilient PUF applications due to their multilevel cell (MLC) characteristics~\cite{reconfPUF,survey20}. 
MLC technology allows NVMs to store multiple bits per cell by using varying voltage levels. This feature does not only enhances the storage density of NVMs, but also contributes to their suitability for PUF implementations. 
The MLC nature introduces additional variability and complexity into the data stored, which can be exploited to create robust and unique PUF responses. 
Consequently, the inherent variability of MLC NVMs enhances the uniqueness and resilience of PUF responses to learning-based attacks.

\subsection{Endurance of NVM-based PUFs}
During the deployment of PUFs, they receive challenges to generate responses to authenticate the device containing the PUF. 
For NVM-based PUFs, this means that several writes and reads are conducted to capture the responses.
This can start to damage the PUF, as NVM cells usually degrade in endurance with the number of writes.

This endurance degradation affects the reliability of NVM-based PUFs by altering their responses. To address these effects, few strategies have been proposed. 
Re-enrollment is one such strategy; it involves periodic updates to the PUF's CRPs via trusted third-party data which could accommodate changes over time~\cite{reconfPUF}. 
Another approach involves adjusting the operational range of the PUFs to address any deviations in response patterns~\cite{iccad_age} to mitigate noise in general, which can include a change of endurance. 
In addition, some solutions adapt techniques from silicon delay-based PUFs to manage reliability over time, applying established methods to ensure consistent performance~\cite{anv-puf,aging_sel}.
However, these methods overlook NVM endurance. 
Thus, an endurance-aware PUF design is timely needed.

\section{NVM-PUF Endurance Analysis}
\label{sec:EndurNVM}

\begin{figure*}[htbp]
  \centering
    \includegraphics[width=0.95\linewidth,page=1]{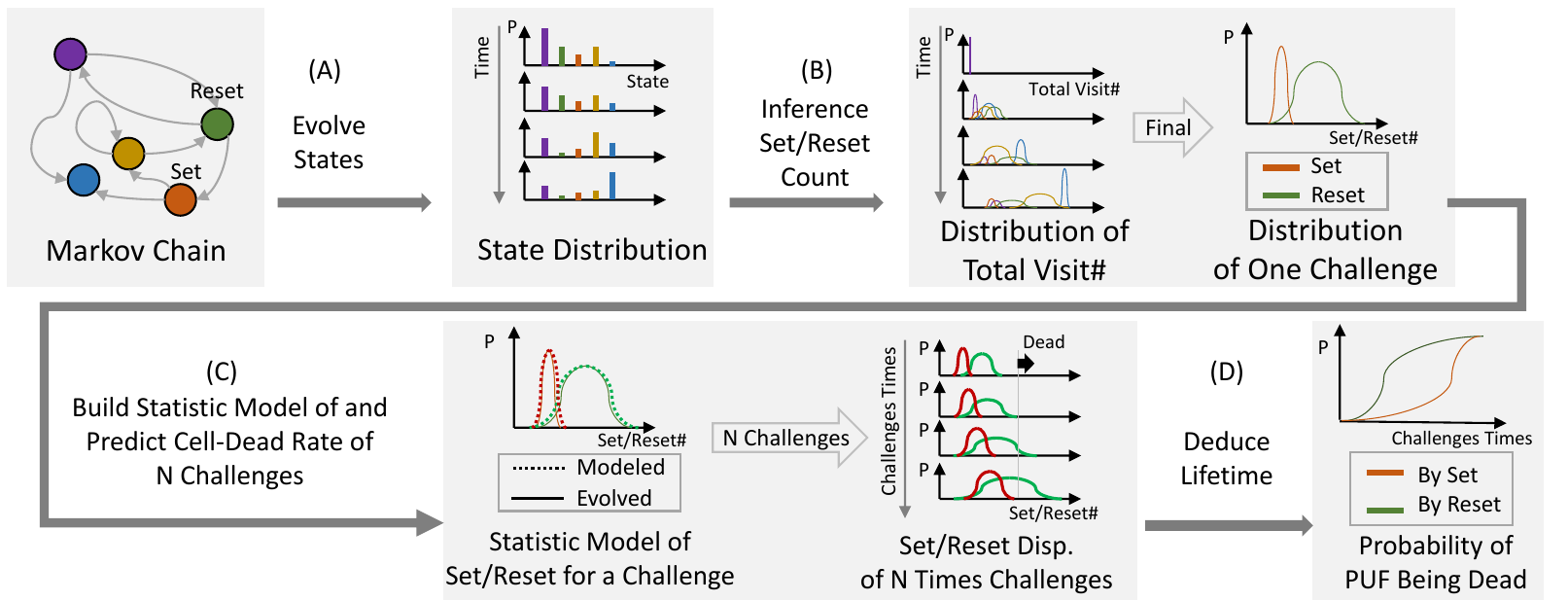}
    \caption{Flow to analyze the endurance of the PUF. Based on a Markov chain reconstruction of the PUF the distribution of set and reset operations can be deduced to infer when would the PUF be unusable.}
    \label{fig:puf_flow}
\end{figure*}

To analyze the lifetime of PUFs, we developed an endurance analysis methodology, shown in \cref{fig:puf_flow}, which uses a Markov chain to deduce the probability that the PUF will fail after $N$ challenges. The detailed explanation is provided in the following subsections.
\subsection{Analyzing States}
\begin{figure}[htbp]
  \centering
    \includegraphics[width=\linewidth,page=2]{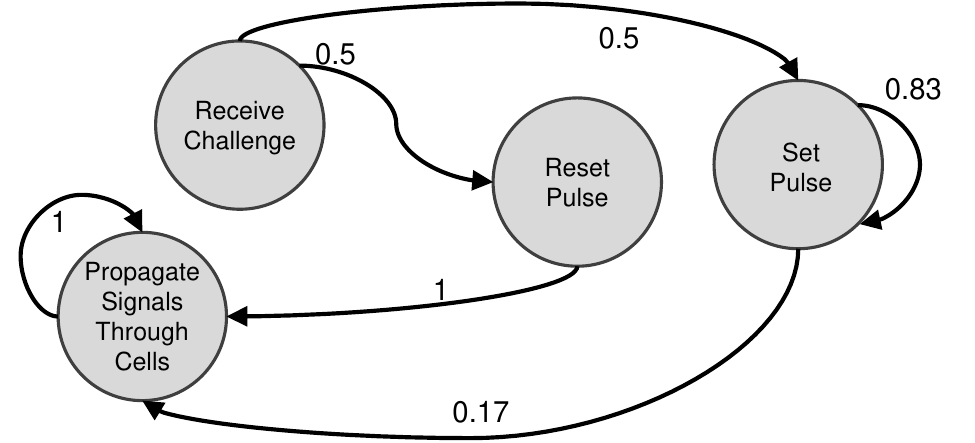}
    \caption{Markov chain modeling the behavior of cells in PUF from~\cite{Arbiter_mpuf}.}
    \label{fig:old_puf_fsm}
\end{figure}

The Markov chain evolves the probability of each state using the transition matrix. The probability of each state is represented as a state vector and the probability at the next moment is determined by applying the state transition matrix to the current state vector. 
All state transitions begin from the ``Receive Challenge'' state. Therefore, the initial state is represented by a one-hot vector, where the probability of the "Receive Challenge" state is one.
As an example, we show the Markov chain of the PUF from~\cite{Arbiter_mpuf} in \cref{fig:old_puf_fsm}.

Additionally, since the state machine includes a terminal state ``Propagate Signals Through Cells'' in \cref{fig:old_puf_fsm},
where the probability of transitioning to other states is zero, 
the system will eventually converge to this terminal state. We set a stop condition for our evolution process when the probability of reaching the termination state is $1 - 10^{-5}$.
Through the iterative update of the system's state vector, we can determine the probability of each state at each iteration.
We denote each state as $P_m(t,s)$, which represents the probability that state s is visited at iteration $t$. These records are then used to calculate the distribution of the accumulated visit time for each state.

\subsection{Inference Set/Reset Count}
\label{set_count}
Since our main concern is the endurance of the PUF system, it is essential to calculate the accumulated set/reset operations on the device. To achieve this, we first need to determine the distribution of the total visit counts in a single challenge for each state. Then, we extract the final distribution of the states associated with the set and reset operations. 

To get the distribution of the total visit count, we utilize the recorded probability of each transition state and convert it to the total visit count by the following equations. 
\begin{equation}
    P_{c}(N|t,s)=P_{c}(N-1|t-1,s)P_m(t-1,s)+KEEP(t,s)
    \label{eq:prob_acc}
\end{equation}
\begin{equation}
    KEEP(t,s)=P_{c}(N|t-1,s)(1-P_m(t-1,s))
    \label{eq:prob_acc_con}
\end{equation}

\begin{equation}
P_{c}(0|t,s)=
\begin{cases}
    1, & \text{if } t = 0 \\
    KEEP(t,s), & \text{otherwise}
\end{cases}
    \label{eq:prob_acc_ini}
\end{equation}

In \cref{eq:prob_acc}, $P_{c}(N|t,s)$ denotes the probability of state $s$ being visited $N$ times by iteration $t$. This probability can be derived from two components: the probability that the visit count has just reached $N$ at iteration $t-1$, as expressed in the first part of \cref{eq:prob_acc}, and the probability that the visit count had already reached $N$ before iteration $t-1$ with no subsequent updates, as described in \cref{eq:prob_acc_con}. The initial condition is provided in \cref{eq:prob_acc_ini}, where the probability of not visiting any state is set to one at the beginning, and the probability of remaining in the non-visiting state is given by $KEEP(s,t)^t$. 

Because the system eventually transitions to the ending state, the states related to set and reset operations converge to being unvisited. In other words, the total number of visits associated with the set and reset operations ceases to update and instead converge to a stable distribution, which we consider as the set and reset distribution for a given challenge.
\subsection{Modeling of Set/Reset Distribution}
We consider the set and reset counts of a challenge as random variables that follow the distribution introduced in \cref{set_count}. Based on this distribution, we deduce the set and reset counts after N challenges. The distribution of the total number of set and reset operations after N challenges is derived by summing N independent random variables, each following the set and reset operation distribution of an individual challenge. Finally, we obtain the time-variant distribution of the set and reset operations, enabling us to deduce the lifetime of the system.
\subsection{Deduce Lifetime}
\label{sec:lifetime_calc}
The cycling endurance of a PCM cell is considered to be 1000 cycles, as reported in prior studies~\cite{anand2021cycle,dheep2015influence,shukla2008thermal,sari2011synthesis,aydin2013fatty,tyagi2008thermal,el2011one}. This endurance limit acts as a threshold; any cell whose set and reset count exceeds this threshold is considered dead. A PUF may contain M cells and is considered dead when 15\% of its cells are dead, which is a common threshold of PUF reliability~\cite{pufatt14,singleround19,anv-puf}. Our goal is to deduce the probability of the PUF failing based on the above definitions and the distribution of set and reset operations.

For each challenge $t$, the probability that a single cell is dead is the probability that the number of set and reset operations exceeds the cell’s endurance limit, as defined in \cref{eq:deadP}. Since each cell operates independently, the distribution of $k$ dead cells among $M$ cells follows a binomial distribution, as expressed in \cref{eq:binominal}. Finally, we calculate the probability of the PUF being dead by adding the probabilities of having k dead cells for $k>0.15M$ to \cref{eq:PUFdead}.
\begin{equation}
    P_{cell}(dead|t)=P(set~or~reset~ops. > limitation)|_{t}
    \label{eq:deadP}
\end{equation}
\begin{equation}
    P(k~dead|t)=C_k^MP_{cell}^k(1-P_{cell})^{M-k}|_{t}
    \label{eq:binominal}
\end{equation}
\begin{equation}
    P(PUF~dead|t)=\Sigma_{k>0.15M}P(k~dead)|_{t}
    \label{eq:PUFdead}
\end{equation}
\section{Endurance-Aware PUF: REAP-NVM}
\label{sec:proTech}

To respect the limited endurance while providing a secure PUF against learning-based attacks, our design, named REAP-NVM, uses the multilevel cell characteristics of NVM to mitigate ML modeling attacks.
Our goal is to have a strong PUF with a large pool of CRPs that can be used effectively for authentication and key generation.
Moreover, we design our REAP-NVM to avoid the endurance degradation.
Instead of writing to all cells after receiving a challenge, only one pair of cells is written, while the others remain unchanged.
Thus, cell aging is greatly reduced and balanced without sacrificing security, as shown in \cref{sec:res}.
This is based on the concept of interpose PUF~\cite{interpose} where changing one challenge bit increases security with a low overhead.

\begin{figure}[htbp]
  \centering
    \includegraphics[width=\linewidth,page=1]{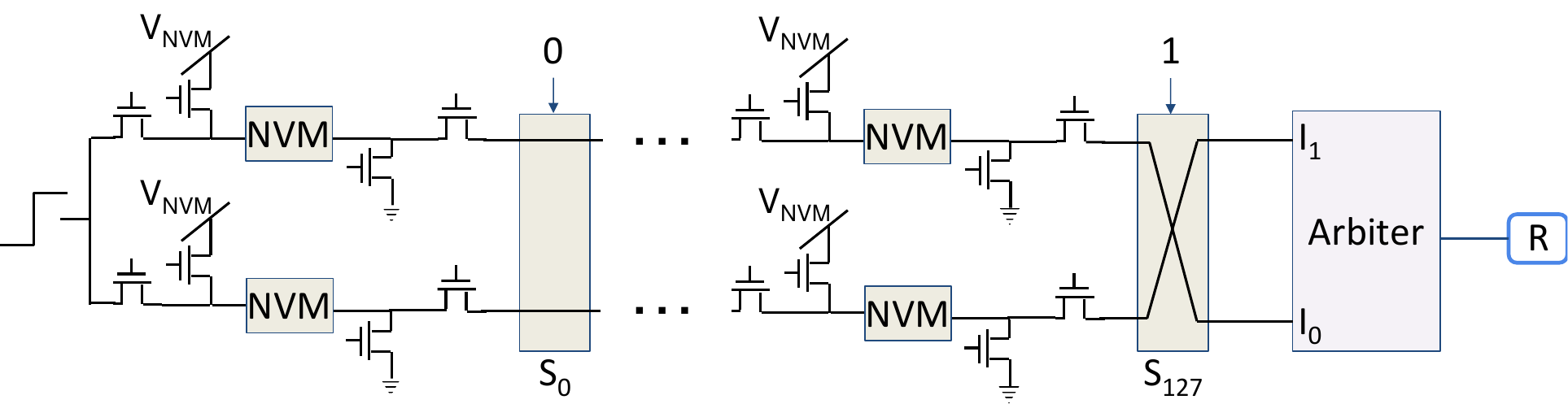}
    \caption{Design of REAP-NVM. Similar to an AUF but with the addition of NVM cells with variable delay to increase the security.}
    \label{fig:puf_des}
\end{figure}

\Cref{fig:puf_des} shows the design of our PUF.
It is based on the Arbiter PUF (APUF) design with 128~stages.
However, instead of wiring the switches directly, we add NVM cells in between them.
Each challenge consists of three parts.
First, a 128\,bit value that controls the individual switches, if the bit is `0', the switch stays in parallel configuration, else, it crosses the paths.
The second part is 7 bits, choosing one pair of NVM cells to be configured to an arbitrary level.
The arbitrary level is set using the third and final part of the challenge which consists of 3\,bits as several NVM technologies support up to 8 levels~\cite{rram_spring2020}.
Each level consists of a different resistance state which affects the delay of a signal propagating through the PUF which makes it harder for the ML models to correctly predict the behavior of REAP-NVM.

To be able to set the NVM cell to a certain level but also to propagate the signals, a network of transistors is used.
Each cell has two transistors before it and two transistors after it.
If the cell must be set to a certain level or reset, the two transistors connected to \texttt{$\text{V}_{\text{NVM}}$} and \texttt{gnd} are enabled. 
\texttt{$\text{V}_{\text{NVM}}$} is a variable voltage that can be adjusted to the set, reset, or read pulse voltage of the NVM.
To propagate the signals and evaluate the response, the transistors that connect the NVM cell to the switches are enabled.
Note that the control of the transistors, the challenges as input to the switches, and the signal to be propagated are all given to the PUF by a control circuit which is omitted from the figure for simplicity.

\begin{figure}[tbhp]
  \centering
    \includegraphics[width=\linewidth,page=1]{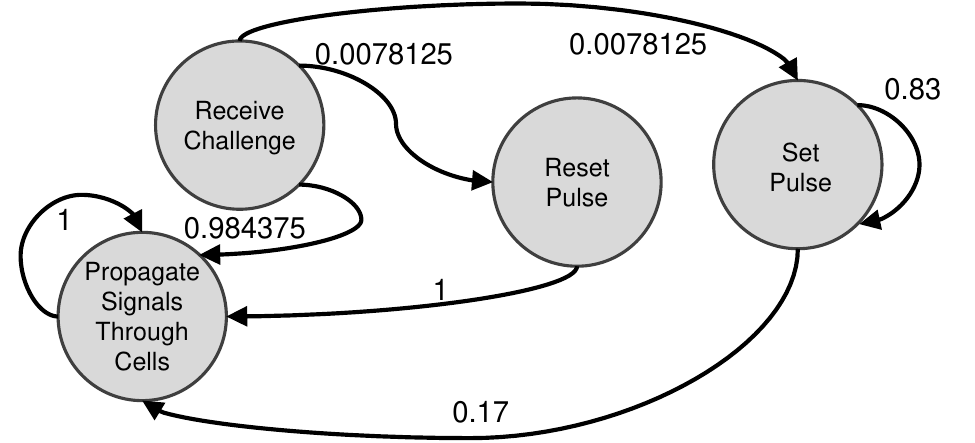}
    \caption{Markov chain modeling the behavior of cells in REAP-NVM. It enhances over the state-of-the-art design from~\cite{Arbiter_mpuf} by reducing the probability of set and reset per cell.}
    \label{fig:puf_fsm}
\end{figure}

To better understand how REAP-NVM is endurance aware, we model its individual cell behaviors as a Makrov chain similar to \cref{sec:EndurNVM}.
\Cref{fig:puf_fsm} shows the Makrov chain; as only one pair of cells is set to an arbitrary level per challenge, a cell has a $\frac{1}{128}$ chance to go to the set loop.
Moreover, a cell also has a $\frac{1}{128}$ chance to have been used in the previous challenge and, therefore, may need to be reset to the default value.
This leaves a chance of $\frac{126}{128}$ that a cell will not be touched at all during the generation of the response from a challenge.
As with each new challenge, a random pair of cells is chosen, the usage of the cells overtime will average out, and no certain cell will be overused, lowering the risk of damaging the PUF.

In addition to balancing cell usage, REAP-NVM enhances security by leveraging the multilevel cell characteristics of NVM. Using multiple resistance levels within NVM cells, each CRP becomes more complex and unpredictable. 
This increased complexity significantly hampers the ability of ML models to accurately predict the PUF's behavior, as the added levels introduce a greater degree of variability and noise into the system. 
Consequently, even if an attacker were to acquire a substantial dataset of CRPs, the inherent unpredictability introduced by multilevel cells would still pose a significant barrier to successful modeling and prediction.

Furthermore, the selective writing strategy employed by REAP-NVM not only mitigates aging, but also reduces power consumption. 
Since only one pair of cells is modified per challenge, the overall energy required for PUF operation is minimized. 
This makes REAP-NVM not only more durable but also more energy-efficient, which is particularly advantageous for resource-constrained applications such as IoT devices. 
The combination of enhanced security, reduced aging effects, and lower power consumption makes REAP-NVM a robust and efficient solution for modern cryptographic applications.

\section{Evaluation}
\label{sec:res}

We first evaluate REAP-NVM to know whether it achieves desirable behavior or not.
We build a SPICE/Matlab simulation environment using PCM and get the model parameters from~\cite{anv-puf}.
We simulate two REAP-NVM PUFs and generate 102,400 CRPs from each of them.
Moreover, we simulate a normal APUF with 128 stages to compare against it as a baseline.

\begin{figure}[tbhp]
  \centering
    \includegraphics[width=\linewidth,page=1]{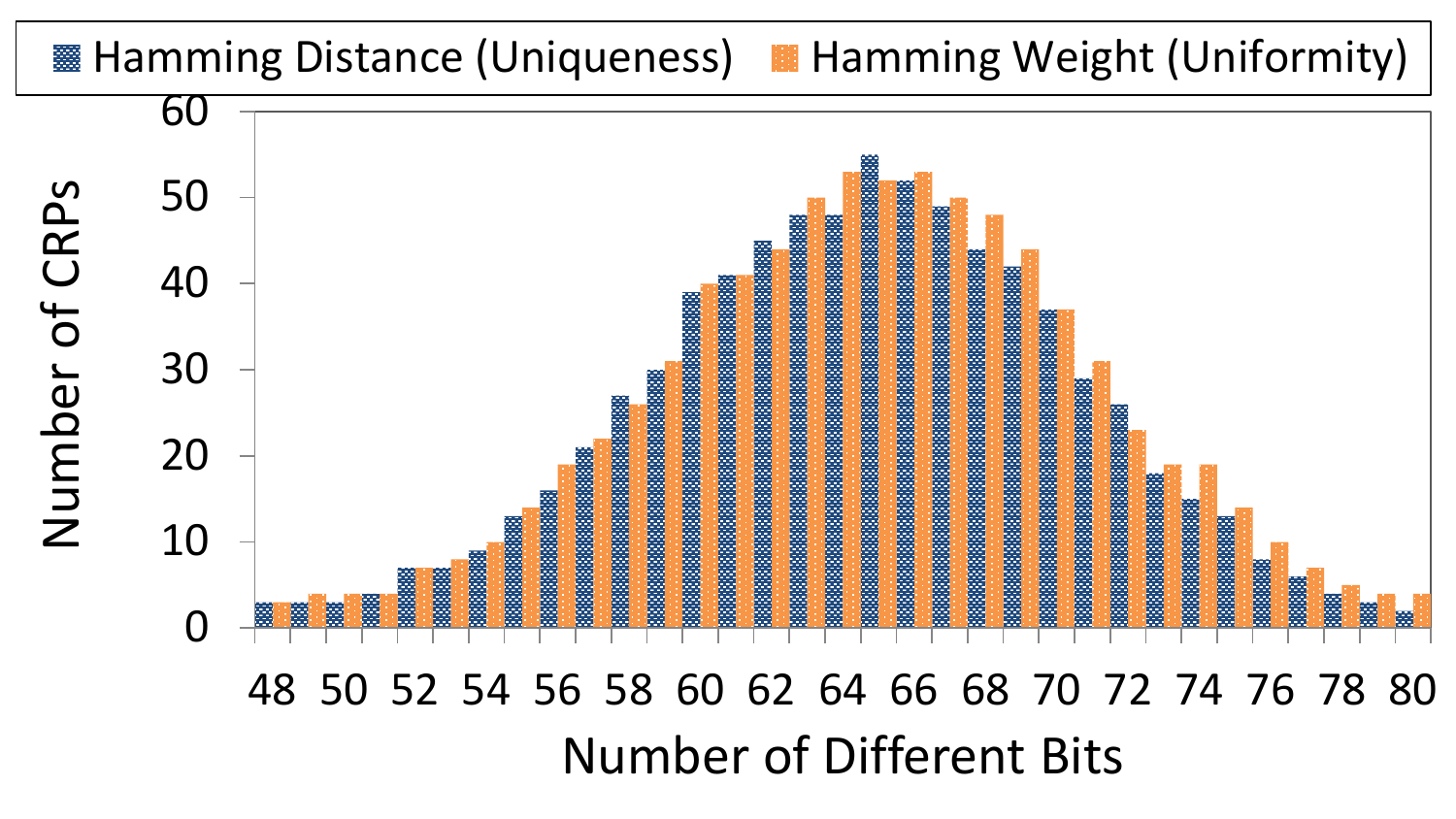}
    \caption{Distribution of REAP-NVM's Uniformity and Uniqueness. Both have the desired Gaussian distribution centered around 64.}
    \label{fig:puf_uni}
\end{figure}

The first metrics we evaluate are the uniqueness and uniformity of REAP-NVM as \cref{fig:puf_uni} shows.
If the uniqueness or uniformity are poor, the PUF will be clonable, and thus it does not even need an ML model to predict its output.
Our responses are natively of 1\,bit length; however, to evaluate uniqueness and uniformity, we assume that each 128 responses will be packed together as one response.
This gives us 800 responses that we can compare their hamming weight per PUF and compare the hamming distances between them.
As \cref{fig:puf_uni} shows, the distribution of both is in the desired bell shape around 64 which is the middle of the response bitwidth.
Therefore, REAP-NVM has good uniqueness and uniformity.

\begin{figure}[htbp]
  \centering
    \includegraphics[width=\linewidth,page=3]{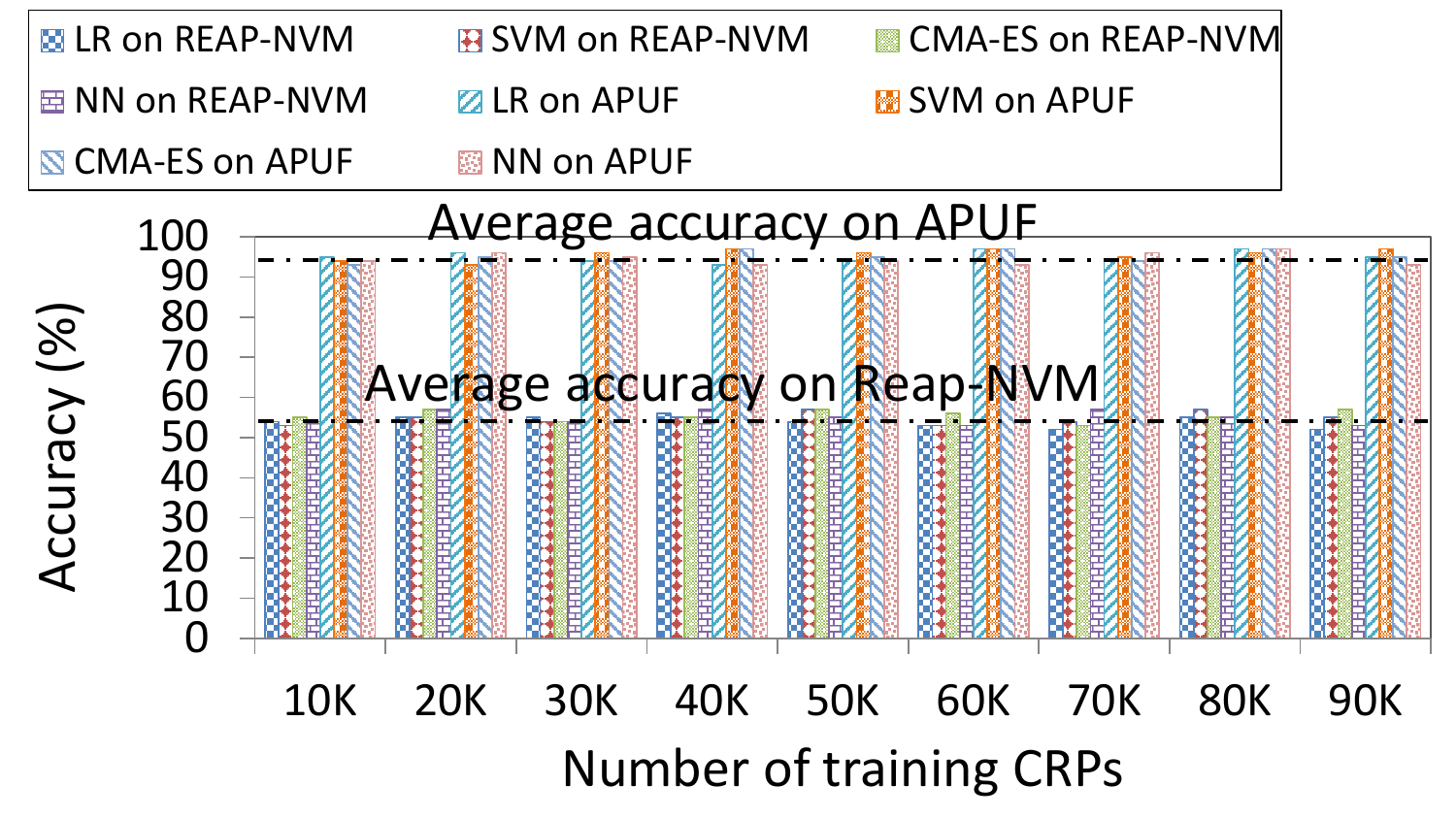}
    \caption{Performance of learning-based attacks on REAP-NVM. Compared to the baseline of APUF, REAP-NVM is much secure.}
    \label{fig:puf_res}
\end{figure}

Next, we evaluate the security of REAP-NVM against learning-based attacks. 
We use the same attacks as those from~\cite{flamPUF}.
We increase the training set from 10,000 to 90,000 CRPs with a step 0f 10,000 CRPs.
As a baseline, we compare the prediction accuracy with APUF.
As \cref{fig:puf_res} shows, REAP-NVM remains resilient to learning-based attacks with a prediction accuracy of around 55\%, that is, in a range similar to flipping a coin.
However, APUF is easily predictable, having already been in the range of 95\% from the lowest training dataset.

 \begin{figure}[htbp]
   \centering
     \includegraphics[width=1\linewidth]{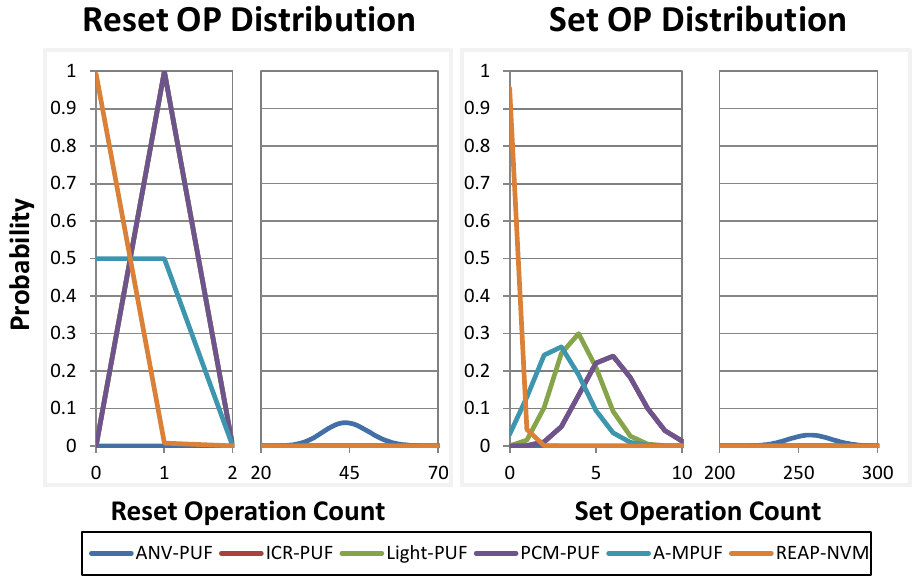}
     \caption{Distribution of Reset and Set Operation for a Challenge, for one challenge a significantly higher distribution of set operations is performed compared to reset operations.}
     \label{fig:set_reset_disp}
 \end{figure}

After ensuring that REAP-NVM is unique, uniform, and mitigates the learning-based attacks, we analyze its endurance in comparison to the state-of-the-art.
In addition to REAP-NVM, we also analyze A-MPUF~\cite{Arbiter_mpuf} shown in \cref{fig:old_puf_fsm} and ICR-PUF~\cite{PCM_PUF_Model14}, Light-PUF~\cite{reram_iterative}, and PCM-PUF~\cite{PCM_iterative}.
The first step is to get the set and reset distribution for each of them after modeling them as Markov chains based on the analysis from \cref{set_count}.
As \cref{fig:set_reset_disp} shows, REAP-NVM has a very low number of set and reset operations compared to the state-of-the-art.
Moreover, as a general trend, the number of set operations dominates the reset operations and would have a higher role in the lifetime of the NVM-based PUFs.

\begin{figure}[htbp]
  \centering
    \includegraphics[width=1\linewidth]{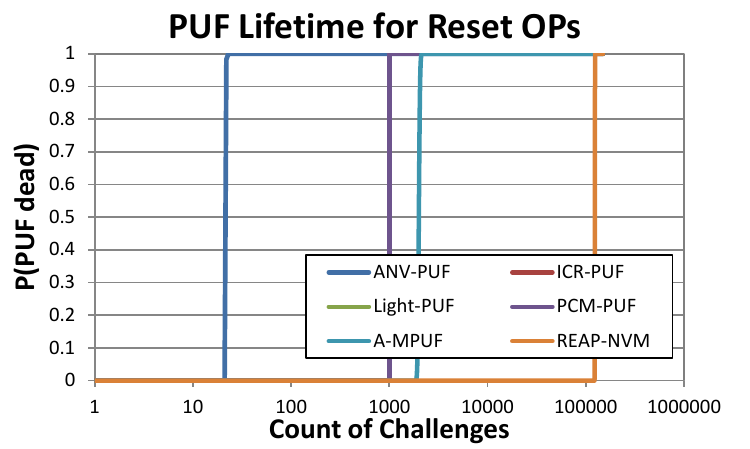}
    \caption{Probability distribution of PUF Lifetime based on set operations.}
    \label{fig:lifetime_reset}
\end{figure}

After getting the distributions, we get the lifetime probability distribution based on the analysis from \cref{sec:lifetime_calc}.
Starting with the reset distribution, \cref{fig:lifetime_reset} shows the probability distribution of REAP-NVM along with the state of the art.
It can be seen that REAP-NVM improves significantly over the state-of-the-art.
Even with a logarithmic X-axis, the probability of having REAP-NVM PUF dead occurs significantly after many more challenges compared to A-MPUF~\cite{Arbiter_mpuf} the next best performing PUF.
Moreover, it can be seen that ANV-PUF~\cite{anv-puf} has the worst lifetime.
It can also be noted that the change in probability is steep for all PUFs from `0' to `1'.

\begin{figure}
  \centering
    \includegraphics[width=1\linewidth]{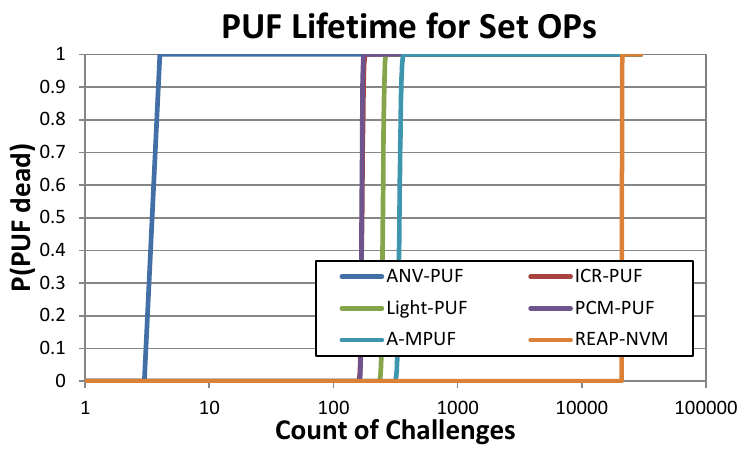}
    \caption{Probability distribution of PUF Lifetime based on set operations.}
    \label{fig:lifetime_set}
\end{figure}

This trend generally continues with the lifetime probability based on set pulses.
REAP-NVM stays best, A-MPUF~\cite{Arbiter_mpuf} second, and ANV-PUF~\cite{anv-puf} worst.
However, the number of challenges is significantly less, i.e., the NVM-based PUFs would be dead based on the set operations not the reset operations.
Moreover, specially pronounced for ANV-PUF~\cite{anv-puf}, the transition is less steep.
Hence, the PUF might be dead earlier than expected.

\begin{figure}[htbp]
  \centering
    \includegraphics[width=1\linewidth]{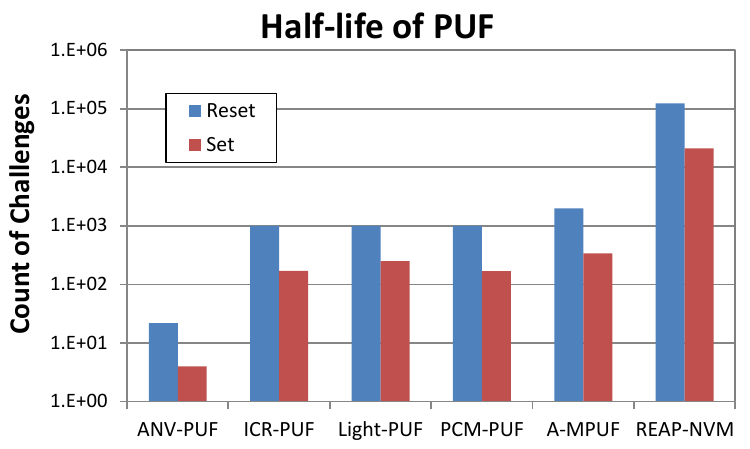}
    \caption{Half-Life of PUF affected by Set and Reset based on the probability distribution of the lifetime. Our proposed REAP-NVM is the best performing.}
    \label{fig:half_life}
\end{figure}

Based on the probabilities of the lifetime for PUFs we evaluate the half-life of each PUF, i.e., when the probability of PUF being dead is 50\%.
\Cref{fig:half_life} shows the half-life evaluation.
ANV-PUF~\cite{anv-puf} has a very low half-life, making it barely usable.
Other state-of-the-art PUFs are better, but not comparable to REAP-NVM.
Overall, compared to the next best PUF, REAP-NVM has $62\times$ improvement.

\begin{table}[htbp]
    \centering
    \caption{Comparison to the related works, based on a table from~\cite{anv-puf}. REAP-NVM has a relatively-low energy consumption and significantly high endurance.}
    \begin{tabular}{cccccc}
    \hline
 	PUF & ML & Strong & Energy & Half-life & Cells per\\
        type & Resil. & PUF & ($J$) & (Set) & Resp. bit\\
 	\hline
 	\textbf{REAP-NVM}& yes & yes & 752\,$n$ & 21099 & 256\\
 	ANV-PUF~\cite{anv-puf} & yes & yes & 7.2\,$\mu$ & 4 & 2\\
 	A-MPUF~\cite{Arbiter_mpuf} & partially & yes & 575\,$n$ & 341 & 256\\
 	PCM-PUF~\cite{PCM_iterative} & yes & no & 1.9\,$\mu$ & 169 & 0.25\\
        Light-PUF~\cite{reram_iterative} & yes & no & 150\,$n$ & 251 & 1\\
        ICR-PUF~\cite{PCM_PUF_Model14} & yes & no & 625\,$n$ & 171 & 1\\
 	\hline
    \end{tabular}
    \label{tab:detailed_results}
\end{table}

Although endurance is our main improvement metric, energy is also an important aspect that we investigate as well. 
\Cref{tab:detailed_results} shows the comparison of energy and other metrics between REAP-NVM and the state-of-the-art PUFs. It is based on a similar table from~\cite{anv-puf}. 
For energy consumption, we use the numbers from~\cite{PCMmodeling21}.
Although REAP-NVM does not have the lowest energy consumption, it is still relatively low in the range of hundreds of~$n J$.
Moreover, when combined with the other metrics, REAP-NVM is the only strong, fully ML-resilient PUF that has an energy consumption in the $n J$~range that is usable from the endurance point of view.
Our PUF's main weakness is the area overhead. It requires 256 NVM cells to produce one response bit.
In contrast, other PUFs can produce one response bit using one cell or even up to 4 bits per cell.
Although this reduces their lifetime, it results in a lower area.

\section{Conclusions}
\label{sec:conc}
Non-Volatile Memory (NVM) technologies have emerged as a promising solution for designing Physical Unclonable Functions (PUFs), against learning-based attacks.
However, a significant issue with NVM-based PUFs is their endurance problem; frequent write operations lead to wear and degradation over time, reducing the reliability and lifespan of the PUF.
In this work, we address these issues by offering an analytical model to predict the degradation of endurance, and investigate various state-of-the-art PUF designs.
We propose a novel endurance-aware PUF design, namely REAP-NVM, which mitigates learning-based attacks while addressing the endurance issues associated with NVM-based PUFs.
The experimental results show that our REAP-NVM achieves a $62\times$ improvement in endurance compared to the state-of-the-art without a reduction in security.
\bibliographystyle{ieeetran_MOD}
\bibliography{references}
\end{document}